\documentstyle[12pt]{article}
\newcommand{\be}{\begin{equation}} \newcommand{\ee}{\end{equation}} \date{}

\begin{document}
\begin{titlepage} \begin{flushright} HD--THEP--94--2\\ \end{flushright}
\vspace{1.8cm} \begin{center} {\bf\LARGE Flow Equations and BRS Invariance}\\
\vspace{.5cm} {\bf\LARGE for Yang-Mills Theories}\\ \vspace{1cm} Ulrich
Ellwanger\footnote{Supported by  a DFG Heisenberg fellowship;  e-mail: I96 at
VM.URZ.UNI-HEIDELBERG.DE}\\ \vspace{.5cm} Institut  f\"ur Theoretische Physik
\\Universit\"at Heidelberg\\ Philosophenweg 16, D-69120 Heidelberg, FRG\\

\vspace{2cm} {\bf Abstract:}\\ \end{center} \parbox[t]{\textwidth}{ Flow
equations describe the evolution of the effective action $\Gamma_k$ in  the
process of varying an infrared cutoff $k$. The presence of the infrared cutoff
explicitly breaks gauge and hence BRS invariance. We derive modified
Slavnov-Taylor identities, which are valid for nonvanishing $k$. They
guarantee the BRS invariance of $\Gamma_k$ for $k\to0$, and hence allow the
study of non-abelian gauge theories by integrating the flow equations.
Within a
perturbative expansion of $\Gamma_k$, we derive an equation for a $k$
dependent
mass term for the gauge fields implied by the modified Slavnov-Taylor
identities.}

\end{titlepage}

The use of flow equations (or exact renormalization group equations \cite{1}
resp. evolution equations) in continuum quantum field theory has recently been
the subject of numerous investigations. They were employed by Polchinski and
many others \cite{2}-\cite{6} in order to simplify proofs of perturbative
renormalizability. Here the flow equations are used to construct bare actions
(depending on an ultraviolet cutoff $\Lambda$), which guarantee the existence
of
finite quantum effective actions in the limit $\Lambda \to \infty$.

Alternatively the flow equations can be used to construct quantum effective
actions in terms of bare actions, with an ultraviolet cutoff $\Lambda$ kept
fixed. Their integration with respect to an infrared cutoff $k$ then
 serves as a computational tool, which allows to calculate the
generating functionals (or the Green functions) in the limit $k\to 0$ in terms
of boundary conditions for those functionals at some large infrared cutoff
$k=\bar k$. The result corresponds to the one of a quantum field theory with
fixed ultraviolet cutoff $\Lambda =\bar k$, with a classical action related to
the boundary conditions at $k=\bar k$. For recent work in this direction see
refs. \cite{7},\cite{8}.

The simple form and the exactness of the flow equations is based
on the fact that the cutoffs are introduced by modifying the propagators of
the
fields. In momentum space, e.g., the propagators get multiplied by a function
of $p^2$, which vanishes rapidly for momenta beyond the cutoffs. The
application
of the  present concept of flow equations to gauge theories leads thus to
serious problems, since the presence of such cutoffs breaks gauge invariance
explicitly. (Kleppe and Woodard \cite{9} studied a closely related
regularization which affects not just the propagators, but modifies the entire
action. This method preserves distorted versions of gauge symmetries, but the
resulting effective action no longer satisfies simple flow equations.)

In \cite{8} background gauge fields were introduced in order to cope with this
situation, but their presence in the final expressions for the generating
functionals leads to new conceptual and practical difficulties. In
\cite{4}-\cite{6} Ward or  Slavnov-Taylor (ST) identities \cite{10} were
employed in order to obtain proofs of perturbative renormalizability for gauge
theories.

Let us denote by $G_k^{\Lambda}(J)$, $\Gamma_k^\Lambda(\phi)$ the generating
functionals of connected and one-particle irreducible Green functions,
respectively. The indices $k$ and $\Lambda$ refer to the fact that only modes
with $k^2<p^2<\Lambda^2$ have been
integrated out, i.e. $G_k^\Lambda$ and $\Gamma_k^\Lambda$ are computed in the
presence of an infrared cutoff $k$ and an ultraviolet cutoff $\Lambda$ in the
propagator. For $k=\Lambda$ $G_\Lambda^\Lambda$ and $\Gamma_\Lambda^\Lambda$
are
simply related to a bare action $S_0(\Lambda)$ \cite{3}, \cite{7}. Proofs of
perturbative renormalizability start with the construction of $G_\Lambda^
\Lambda$
resp. $\Gamma_\Lambda^\Lambda$ and hence $S_0(\Lambda)$, given the knowledge of
the relevant or marginal couplings of $G_0^\Lambda$ and $\Gamma_0^\Lambda$, by
integrating the flow equations for $G_k^\Lambda$ resp. $\Gamma_k^\Lambda$.
Subsequently the limit $\Lambda \to \infty$ has to be shown to exist, keeping
$G_0^\Lambda$ and $\Gamma_0^\Lambda$ finite. In \cite{4}-\cite{6} Ward or
Slavnov-Taylor (ST) identities were imposed on the relevant or marginal
couplings of $G_0^\Lambda$ resp. $\Gamma_0^\Lambda$, and it was argued, that a
bare action $S_0(\Lambda)$ consistent with these identities can be constructed.

This procedure is of no help, however, if we want to compute the generating
functionals for $k \to 0$ in terms of $S_0(\Lambda)$; we do not know the
relevant or marginal couplings of $G_0^\Lambda$ resp. $\Gamma_0^\Lambda$
beforehand. It should be clear that $G_\Lambda^\Lambda$
resp. $\Gamma_\Lambda^\Lambda$ and hence $S_0(\Lambda)$ cannot satisfy the
standard ST identies, which are related to BRS invariance \cite{11} and hence
to
gauge symmetry. We need a condition on the breaking on BRS symmetry of
$G_\Lambda^\Lambda$ resp. $\Gamma_\Lambda^\Lambda$ and hence $S_0(\Lambda)$,
which compensates precisely the breaking of BRS symmetry induced by the
integration over modes with a finite cutoff $\Lambda$. Only such a form of
$S_0(\Lambda)$ can guarantee the validity of the standard ST identities for
$G_0^\Lambda$ resp. $\Gamma_0^\Lambda$ and hence the unitarity of the theory.

Below we derive such a condition in the form of modified ST identities for
$G_k^\Lambda$ resp. $\Gamma_k^\Lambda$, which enjoy the following properties:

1. If they are satisfied for some particular value of $k$, e.g. for $k=\bar k$,
they are automatically satisfied for any other value of $k$, provided
$G_k^\Lambda$ and $\Gamma_k^\Lambda$ are related to $G_{\bar k}^\Lambda$
and $\Gamma_{\bar k}^\Lambda$ by
integrating the flow equations.

2. In the limit $k\to0$ they turn into the standard ST identities.\\ Hence, if
we require the boundary conditions $G_{\bar k}^\Lambda$ or $
\Gamma_{\bar k}^\Lambda$ to satisfy
the modified ST identities, the BRS invariance of $G_{k=0}^\Lambda$ and
$\Gamma_{k=0}^\Lambda$ is guaranteed.

In fact these modified ST identities are closely related to the "fine tuning
condition" for the effective Lagrangian derived and studied by Becchi \cite{5}.
There the fine-tuning condition was analyzed with the aim to construct gauge
symmetry violating counter terms, which compensate precisely for the breaking
of
gauge symmetry induced by the
integration over modes with a finite cutoff $\Lambda$, and solutions were shown
to exist.

Below we will construct a perturbative solution of the modified ST identities.
We will be concerned with the necessary presence of a mass term for the gauge
fields within $\Gamma_k$, for a pure $SU(N)$ Yang Mills theory, for $k \neq 0$
due to the modified ST identities.

A priori we could just present the flow equations together with the form of
the modified ST identities, and show that they satisfy both conditions 1. and
2. above. Since the perturbative solutions of the flow equations have been
shown to correspond to the standard perturbative expansion within various
contexts [3] - [8], this constitutes a self-consistent definition of the
theory at the perturbative level.

Instead of this ad hoc presentation we prefer to motivate the
explicit form of the modified ST identities in a way which, however,
implicitely assumes the existence of an invariant UV regularisation. Its
existence might be doubted; this is irrelevant, however, since we do not have
to rely on it in order to define the theory. The final properties of the flow
equations and the modified ST identities are independent of the form of the
derivation and the underlying assumptions.

Let us first specify our conventions for a classical Yang-Mills theory.
(Throughout this paper we work in Euclidean space with arbitrary dimension
$d$.) The Yang-Mills action is given by \be\label{1} S_{YM}=\frac{1}{4}\int
d^dxF^{\ a}_{\mu\nu}F^{\ a}_{\mu\nu}\ee with \be\label{2} F^{\
a}_{\mu\nu}=\partial_\mu A^{\ a}_\nu-\partial_\nu A^{\ a}_\mu+gf^a_{\ bc} A^{\
b}_\mu A^c_\nu.\ee The gauge-fixing part reads \be\label{3} S_{\rm gf}
=\frac{1}{2\alpha}\int d^dx\partial_\mu A^{\ a}_\mu\partial_\nu A^{\ a}_\nu \ee
and the ghost action is given by \be\label{4} S_{\rm gh}=\int
d^dx\partial_\mu\bar c^a(\delta_{ac}\partial_\mu+gf^a_{\ bc}A^{\ b}_\mu) c^c.
\ee
In order to describe the BRS transformations it is convenient to introduce
operators $O^a_{A,\mu}$ and $O^a_g$, which are related to the BRS variations of
$A^a_\mu$ and $c^a$, respectively: \be\label{5} O^{\ a}_{A,\mu}=\partial_\mu
c^a+gf^a_{\ bc}A^b_\mu c^c,\ee \be\label{6} O^a_g=\frac{1}{2}gf^a_{\
bc}c^bc^c.\ee Note that the ghosts $c,\bar c$ and hence the operator $O_A$ are
Grassmann-valued. The sum $S_{YM}+S_{\rm gf}+S_{\rm gh}$ is not
gauge-invariant,
but invariant under BRS transformations of $A^a_\mu,c^a$ and $\bar c^a$
employing a Grassmann parameter $\zeta$: \begin{eqnarray}\label{7} &&\delta
A^{\ a}_\mu=O^{\ a}_{A,\mu}\zeta,\nonumber\\ &&\delta c^a=O^{\
a}_g\zeta,\nonumber\\ &&\delta\bar c^a=-\frac{1}{\alpha}\partial_\mu
A_\mu\zeta.\end{eqnarray} Note that the operators $O_A$ and $O_g$ are
BRS-invariant as well.

The standard definition of the generating functional $G_k$ of connected Green
functions makes use of sources $J_\mu^{\ a},  \chi^a$ and $\bar\chi^a$
for the fields $A_\mu^{\ a},\bar c^a$ and $c^a$, respectively, and additional
sources  $K^{\ a}_\mu$ and $L^a$ coupled to the operators $O^a_{A,\mu}$ and
$O^a_g$. (The sources $\chi^a,\bar\chi^a$ and $K^a_\mu$ will be
Grassmann-valued.) We also introduce a term $\Delta S_k$, which will generate
an infrared cutoff both for the gauge fields and the ghosts. The expression for
$G_k$ then becomes \be\label{8} e^{-G_k(J,\chi,\bar\chi,K,L)}= \int {\cal
D}_{Reg}(A,c,\bar c)e^{-(S_{YM}+S_{\rm gf}+S_{\rm gh}+\Delta S_k+S_{c.t.})
+J\cdot A+\bar\chi\cdot c+\chi\cdot\bar c+K\cdot O_A+L\cdot O_g}\nonumber\\ \ee
Here we have assumed the existence of an ultraviolet regularization (indicated
by the index ''$Reg$'' attached to the path integral measure) and counter terms
(present in $S_{c.t.}$), which should be a) invariant under BRS
transformations, b) independent of the infrared cutoff $k$. As mentioned
above, an ultraviolet regularisation with these properties might not even
exist; below, however, we will just require a definition of the functionals of
the theory in terms of solutions of the flow equations rather than the
existence of the path integral (8).

{}From now on we prefer to work in momentum space; expressions like $J\cdot A$
etc are to be read as \be\label{9} J\cdot A\equiv\int DpJ^{\ a}_\mu(-p)A^{\
a}_\mu(p);\quad Dp\equiv \frac{d^dp}{(2\pi)^d}. \ee

The explicit form of the infrared cutoff term $\Delta S_k$ is given by
\be\label{10} \Delta S_k=\int
Dp\left[\frac{1}{2}A^a_\mu(-p)R^k_{\mu\nu}(p^2)A^a_\nu(p)+\bar c^a(-p) \tilde
R^k(p^2)c^a(p)\right].\ee The functions $R^k_{\mu\nu}(p^2)$ (which will
depend on
the gauge parameter $\alpha$) and $\tilde R^k(p^2)$ modify the gauge field and
ghost propagators such that the propagation of modes with $p^2<k^2$ is
suppressed. Convenient choices are \begin{eqnarray}\label{11}
R^k_{\mu\nu}(p^2)&=&(p^2\delta_{\mu\nu}+(\frac{1}{\alpha}-1)p_\mu
p_\nu)R^k(p^2),\nonumber\\ \tilde R^k(p^2)&=&p^2R^k(p^2),\end{eqnarray} where
$R^k(p^2)$ vanishes for $p^2\gg k^2$ and diverges for $p^2\ll k^2$.  With
(\ref{11}) the gauge field propagator $P_{A,\mu\nu}$ and the ghost propagator
$P_g$ become \begin{eqnarray}\label{12}
P_{A,\mu\nu}(p)&=&\left[\frac{\delta_{\mu\nu}}{p^2}+(\alpha-1)\frac{p_\mu
p_\nu}{p^4} \right]\cdot\frac{1}{1+R^k(p^2)},\nonumber\\
P_g(p)&=&\frac{1}{p^2}\cdot\frac{1}{1+R^k(p^2)}.\end{eqnarray} A possible
explicit form for $R^k(p^2)$ is given by \be\label{13}
R^k(p^2)=\frac{e^{-(p^2/k^2)}}{1-e^{-(p^2/k^2)}}.\ee Now we use the
representation eq. (\ref{8}) for $G_k$ in order to motivate the corresponding
flow equations. We simply have to differentiate both sides with respect to $k$.
On the r.h.s. we obtain an expectation value of $(-\partial_k \Delta S_k$),
which is quadratic in the fields according to eq. (\ref{10}). After replacing
the fields by variations with respect to the corresponding sources, which
allows to pull the expressions out of the path integral, the flow equation for
$G_k(J,\chi,\bar\chi,K,L)$ becomes \begin{eqnarray}\label{14} \partial_k
G_k&=&\int Dp\Bigl\lbrace\frac{1}{2}\partial_k R^k_{\mu\nu}(p^2)\left[
\frac{\delta G_k}{\delta J^a_\mu(p)}\frac{\delta G_k}{\delta J^a_\nu(-p)}-
\frac{\delta^2G_k}{\delta J^a_\mu(p)\delta J_\nu^a(-p)}\right]\nonumber\\
&&+\partial_k\tilde R^k(p^2)\left[\frac{\delta G_k}{\delta\chi^a(p)}
\frac{\delta
G_k}
{\delta\bar\chi^a(-p)}-\frac{\delta^2G_k}{\delta\chi^a(p)\delta
\bar\chi^a(-p)}\right]
\Bigr\rbrace.\end{eqnarray} Next we perform a Legendre transformation in order
to arrive at the effective action $\Gamma_k(A,c,\bar c,K,L)$. Here $A,c$ and
$\bar c$ denote classical fields, and the sources $K$ and $L$ are the same as
before: \begin{eqnarray}\label{15} G_k(J,\chi,\bar\chi,K,L)&=&\Gamma_k(A,c,
\bar
c,K,L)-(J\cdot A+\bar\chi\cdot c+\chi\cdot \bar c),\nonumber\\ A^{\
a}_\mu&=&-\frac{\delta G_k}{\delta J_\mu^{\ a}}, \quad c^a=-\frac{\delta
G_k}{\delta\bar \chi^a},\quad \bar c^a=-\frac{\delta G_k}{\delta
\chi^a},\nonumber\\ J^{\ a}_\mu&=&\frac{\delta \Gamma_k}{\delta A_\mu^{\
a}},\quad \bar\chi^a= -\frac{\delta \Gamma}{\delta  c^a},
\quad\chi^a=-\frac{\delta \Gamma}{\delta \bar c^a}.\end{eqnarray} After the
Legendre transformation the second variations of $G_k$ with respect to sources,
as present in eq. (\ref{14}), can be expressed in terms of $\Gamma_k$.
Essentially they are given by the inverse of the second variations of
$\Gamma_k$
with respect to the fields; the inverse has to be formed, however, in the
enlarged space spanned by the three fields $\lbrace A,c,\bar c\rbrace$. The
precise  relations, including the correct signs, can most easily be obtained by
varying the second line of eq. (\ref{15}) with respect to the three fields, and
by expressing the variations of $G_k$ with respect to the fields in terms of
variations with  respect to the sources using the third line of eq. (\ref{15}).
It becomes convenient to define \be\label{16}
\frac{\delta^2\Gamma_k^{-1}}{\delta\varphi^a\delta\varphi^b}\equiv-
\frac{\delta^2G_k}{\delta j^a\delta j^b}\ee where $\varphi^a$ denotes $\lbrace
A_\mu^{\ a},c^a,\bar c^a\rbrace$ in the cases where $j^a$ is given by $\lbrace
J_\mu^{\ a},\chi^a,\bar\chi^a\rbrace$, respectively. (Note the association of
$c^a$ with $\chi^a$ and not $\bar\chi^a$; below it will be useful to define
$\bar\varphi^a=\lbrace A_\mu^{\ a},\bar c^a,c^a\rbrace$ in terms of
$\varphi^a=\lbrace A^{\ a}_\mu,c^a,\bar c^a\rbrace$.) Eq. (\ref{16}) is to be
read as a definition of the l.h.s. in terms of the r.h.s.; the l.h.s. is not
to
be identified with $(\delta^2\Gamma_k/\delta\varphi^a\delta\varphi^b) ^{-1}$.
Now the flow equations for $\Gamma_k(A,c,\bar c,K,L)$ can be expressed as
\begin{eqnarray}\label{17} \partial_k\Gamma_k&=&\int
Dp\Bigl\lbrace\frac{1}{2}\partial_kR^k_{\mu\nu}
(p^2)\left[A^a_\mu(-p)A^a_\nu(p)+\frac{\delta^2\Gamma^{-1}_k}{\delta A^a_\mu
(p)\delta A^a_\nu(-p)}\right]\nonumber\\ &&+\partial_k\tilde R^k(p^2)\left[\bar
c^a(-p)c^a(p)-\frac{\delta^2\Gamma_k^{-1}} {\delta\bar
c^a(p)\delta^ac(-p)}\right]\Bigr\rbrace.\end{eqnarray} It assumes an even
simpler form, if we introduce $\hat\Gamma_k$ by \be\label{18}
\Gamma_k=\hat\Gamma_k+\Delta S_k;\ee the flow equation for $\hat\Gamma_k$
simply reads \be\label{19} \partial_k\hat\Gamma_k=\int
Dp\left\lbrace\frac{1}{2}\partial_kR^k_{\mu\nu}
\frac{\delta^2\Gamma^{-1}_k}{\delta A^a_\mu(p)\delta A^a_\nu(-p)}
-\partial_k\tilde R^k\frac{\delta^2\Gamma^{-1}_k}{\delta\bar c^a(p)
\delta c^a(-p)}
\right\rbrace.\ee

Next we turn to the modified ST identities. The starting point
is again the path integral representation eq. (\ref{8}) for $G_k$. We perform a
field redefinition of the three fields $A\to A+\delta A,\ c\to c+\delta c,\
\bar c \to\bar c+\delta c$, where the variations have the form of BRS
transformations (\ref{7}). Since this field redefinition does not affect the
complete expression, we obtain a condition for the vanishing of a sum of
expectation values, which are not manifestly BRS invariant: \be\label{20}
\langle -\delta\Delta S_k+J\delta A+\bar\chi\delta c+\chi\delta\bar c\rangle=0.
\ee After replacing fields by variations with respect to sources as before, and
$\delta A, \delta c$ by variations with respect to $K,L$, we obtain the
following identity for $G_k$:
\begin{eqnarray}\label{21} &&\int Dp\Bigl\lbrace
J^{\ a}_\mu(p)\frac{\delta G_k}{\delta K_\mu^{\ a}(p)}
+\bar\chi^a(p)\frac{\delta G_k}{\delta L^a(p)}+\frac{i}{\alpha}
p_\mu\chi^a(p)\frac{\delta G_k}{\delta J^{\ a}_\mu(p)}\nonumber\\
&&+R^k_{\mu\nu}(p^2)\left[\frac{\delta G_k}{\delta K_\nu^{\ a}(-p)}\frac{\delta
G_k} {\delta J^{\ a}_\mu(p)}-\frac{\delta^2G_k}{\delta K^{\ a}_\nu(-p)\delta
J_\mu^{\ a} (p)}\right]\nonumber\\ &&+\tilde R^k(p^2)\left[\frac{\delta
G_k}{\delta L^a(-p)}\frac{\delta G_k}{\delta\chi^a
(p)}-\frac{\delta^2G_k}{\delta L^a(-p)\delta\chi^a(p)}\right]\nonumber\\
&&+\frac{i}{\alpha}p_\mu\tilde R^k(p^2)\left[\frac{\delta G_k}{\delta J^{\
a}_\mu(-p)}\frac{\delta G_k}{\delta \bar\chi^a(p)}-\frac{\delta^2G_k}{\delta
J^{\ a}_\mu(-p)\delta\bar\chi^a(p)} \right]\Bigr\rbrace=0.\end{eqnarray} The
first line of eq. (\ref{21}) corresponds to the standard ST identity; the
additional terms involving $R^k_{\mu\nu}$ or $\tilde R^k$ are new and originate
from the term $-\delta\Delta S_k$ in eq. (\ref{20}).

Note that, because of the properties of the functions $R^k_{\mu\nu}$ and $
\tilde
R^k$ (cf. eqs. (\ref{11}) and (\ref{13})), the corresponding momentum  integral
$Dp$ is bound to be ultraviolet-finite. Furthermore, for $k\to0$, the functions
$R^k_{\mu\nu}$ and $\tilde R^k$ vanish, and (\ref{21}) turns into the standard
ST
identity. The equation involves just a one-loop momentum integral $Dp$;
this integration is free of infrared divergencies for non-exceptional
(euclidean) external momenta.
It will be helpful to assign a name to the functional of the  sources
appearing on the l.h.s. of eq. (\ref{21}); we will denote the entire l.h.s. of
eq. (\ref{21}) by $\Sigma_k$.

Next we will have a look at the $k$-dependence of $\Sigma_k$. In evaluating
$\partial_k\Sigma_k$, the $k$-derivative hits $R^k_{\mu\nu},\tilde R^k$ and
$G_k$
present in $\Sigma_k$; the $k$-derivative of $G_k$ is then replaced by the
r.h.s. of the flow equation (\ref{14}). After a straightforward, though
lengthy calculation, the result can be written as
\begin{eqnarray}\label{22}
&&\partial_k\Sigma_k=\int Dp\Bigl\lbrace\partial_kR^k_{\mu\nu}(p^2)\left[
\frac{\delta G_k}{\delta J^{\ a}_\mu(-p)}\frac{\delta\Sigma_k}{\delta J_\nu^{\
a}(p)} -\frac{1}{2}\frac{\delta^2\Sigma_k}{\delta J^{\ a}_\mu(-p)\delta
J_\nu^{\ a}(p)}\right] \nonumber\\ &&+\partial_k\tilde R^k(p^2)\left[\frac
{\delta
G_k}{\delta \chi^a(-p)} \frac{\delta
\Sigma_k}{\delta\bar\chi^a(p)}-\frac{\delta G_k}{\delta\bar\chi^a (-p)}
\frac{\delta \Sigma_k}{\delta \chi^a(p)}-\frac{\delta^2\Sigma_k}
{\delta\chi^a(-p)\delta \bar\chi^a(p)}\right]\Bigr\rbrace.\end{eqnarray} Since
the r.h.s. of (\ref{22}) is linear in $\Sigma_k$, the following important
statement holds: If $\Sigma_k$ vanishes for some $k=\bar k$ (identically in the
sources), it will vanish for all $k$, provided $G_k$ and $G_{\bar k}$ are
related by the integration of the flow equation (\ref{14}). This allows us to
forget about the derivation of eq. (\ref{21}), and to start anew as follows:

Let us define $G_k$ by a functional satisfying the flow equation (\ref{14}).
Its path integral representation (\ref{8}) is helpful for its physical
interpretation, but not required for its definition. Let us now require that,
at some starting point $k=\bar k$, $G_{\bar k}$ satisfies the modified ST
identity (\ref{21}), without referring to its derivation. Eq. (\ref{22}) then
ensures us that $G_k$ will satisfy eq. (\ref{21}) for all $k$, in particular
for $k\to0$. Since eq. (\ref{21}) turns into the standard ST identity for
$k\to0$, we thus obtain the desired property of $G_{k\to0}$. The assumptions
underlying the existence of the path integral representation (\ref{8}) are  not
required any more.

Before we turn to the effective action, we observe that another useful identity
can be derived in a similar fashion: If we perform a redefintion of the field
$\bar c^a$ alone, and use that $\delta S_{\rm gh}/\delta\bar c^a\sim -
\partial_\mu O_{A,\mu}^{\ a}$, we find a relation of the form
\be\label{23}
ip_\mu\frac{\delta G_k}{\delta K_\mu^{\ a}(p)}-\chi^a(-p)+\tilde R^k(p^2)
\frac{\delta  G_k}{\delta\bar\chi^a(p)}=0.\ee

Also this identity can be shown to be invariant under the RG flow.
The form of the modified ST identity for $\Gamma_k$ can simply be obtained by
inserting the Legendre transformations (\ref{15}) into eq. (\ref{21}), as in
the case of the flow equations. We just need two more relations, in order to
express the variations $\delta^2G_k/\delta K\delta J$ and $\delta^2G_k/\delta L
\delta\chi$ in terms of $\Gamma_k$. These have the form
\begin{eqnarray}\label{24}
\frac{\delta^2G_k}{\delta K^{\ a}_\nu(-p)\delta
J_\mu^{\ a}(p)} &=&\int Dq\sum_b\frac{\delta^2\Gamma_k^{-1}}{\delta A^{\
a}_\mu(p)\delta
\varphi^b(q)}\frac{\delta^2\Gamma_k}{\delta\bar\varphi^b(-q)\delta K^a_\nu
(-p)}\nonumber\\ &&\equiv\sum_{\varphi^b}\frac{\delta^2\Gamma^{-1}_k}{\delta
A_\mu^{\ a}(p)
\delta\varphi^b}\frac{\delta^2\Gamma_k}{\delta\bar\varphi^b\delta K_\nu^{\ a}
(-p)},\end{eqnarray} and a similar equation with $K_\nu^{\ a}$ replaced $L^a$.
$\delta^2\Gamma_k^{-1}/ \delta A\delta\varphi^b$ and $\bar\varphi^b$ have been
defined in and below eq. (\ref{16}), respectively.

The ultimate form of the modified ST identity for the effective action becomes
again simpler, if we express it in terms of $\hat\Gamma_k$ defined in eq.
(\ref{18}). It finally reads
\begin{eqnarray}\label{25} &&\int
Dp\Bigl\lbrace\frac{\delta\hat\Gamma_k}{\delta A_\mu^{\ a}(p)}
\frac{\delta\hat\Gamma_k}{\delta K_\mu^{\ a}(-p)}-
\frac{\delta\hat\Gamma_k}{\delta c^a(p)} \frac{\delta\hat\Gamma_k}{\delta
L^a(-p)}-\frac{i}{\alpha}p_\mu A_\mu^{\ a}(p) \frac{\delta\hat\Gamma_k}{\delta
\bar c^a(p)}\nonumber\\
&&-\sum_{\varphi^b}\Bigl[R^k_{\mu\nu}(p^2)
\frac{\delta^2\Gamma_k^{-1}}{\delta A_\mu^{\ a}(p)\delta\varphi^b}
\frac{\delta^2\hat\Gamma_k}{\delta\bar\varphi^b\delta K^{\ a}_\nu(-p)} +\tilde
R^k(p^2) \frac{\delta^2\Gamma_k^{-1}}{\delta c^a(p)\delta\varphi^b}
\frac{\delta^2\hat\Gamma_k}{\delta\bar\varphi^b\delta
L^{a}(-p)}\Bigr]\nonumber\\ &&-\frac{i}{\alpha}p_\mu\tilde
R^k(p^2)\frac{\delta^2\Gamma^{-1}_k} {\delta A^{\ a}_\mu(p)\delta\bar
c^a(-p)}\Bigr\rbrace=0\end{eqnarray} The relation (\ref{23}) becomes in terms
of $\hat\Gamma_k$  \be\label{26} ip_\mu\frac{\delta\hat\Gamma_k}{\delta
K_\mu^{\ a}(p)}+ \frac{\delta\hat\Gamma_k}{\delta \bar c^a(p)}=0\ee and is thus
not at all affected by the infrared cutoff $k$.

We will now present a consistency check of the formalism within lowest order
in perturbation theory. As mentioned above we will concentrate on the mass
term for the gauge fields within $\Gamma_k$.

Within lowest order perturbation theory the flow equation (19) is easily
integrated with the result
\begin{eqnarray}\label{27}  &\hat\Gamma_k = \hat\Gamma_0&+\frac{1}{2} Tr
\Bigl[\ln\left(\frac{\delta^2\hat\Gamma_0}
{\delta A^a_\mu(p)\delta A^b_\nu(q)} + R^k_{\mu\nu}\right)-\ln\left(\frac
{\delta^2\hat\Gamma_0}
{\delta A^a_\mu(p)\delta A^b_\nu(q)}\right)\Bigr] \nonumber\\
&& - Tr \Bigl[\ln\left(\frac{\delta^2\hat\Gamma_0}
{\delta \bar c^a(p)\delta c^b(q)} + \tilde R^k\right)-\ln\left(\frac
{\delta^2\hat\Gamma_0}
{\delta \bar c^a(p)\delta c^b(q)}\right)\Bigr].\end{eqnarray}
Here $Tr$ denotes the trace over Lorentz and gauge group indices as well as
integration over internal momenta with the measure $Dp$ of eq. (9).
Furthermore we made use of $R^{k=0}_{\mu\nu}=\tilde R^{k=0}=0$.

Let us now make an ansatz for $\hat \Gamma_0$ in the form of a classical Yang
Mills action as given by eqs. (1)-(6), $\hat \Gamma_0 =
S_{YM}+S_{gf}+S_{gh}-K\cdot O_A-L\cdot O_g$. After inserting this ansatz for
$\hat \Gamma_0$ into the r.h.s. of eq. (27), an expansion of eq. (27) up to
second order in the gauge fields $A^a_\mu(p)$ and to zeroth order in the
corresponding momentum $p$, one finds that $\hat \Gamma_k$ has to contain a
mass term for the gauge fields of the form $\frac{1}{2}m^2 A^a_\mu A^a_\mu$.
Three diagrams with two external gauge fields contribute to $m^2$: A ghost
loop, a gauge field loop with one quartic vertex and a gauge field loop with
two cubic vertices. The result for $m^2$ reads in $d=4$ dimensions, and with
the form (13) of the the cutoff function $R(p^2)$:
\be\label{28}
m^2=\frac{3Ng^2}{128\pi^2}k^2(\alpha -1)\ee
where $\alpha$ denotes the gauge parameter.

Likewise, the modified ST identity (25) can be solved within a perturbative
expansion. To one loop order, $\Gamma_k$ or $\hat \Gamma_k$ can be replaced by
$\Gamma_0$ resp. $\hat \Gamma_0$ in all terms appearing with factors of
$R^k_{\mu\nu}$ or $\tilde R^k$. After an expansion of eq. (25) to first order
in the gauge field $A^a_\mu(p)$, to first order in the ghost field $c^a(-p)$
and to first order in the momentum $p_\mu$ only the first term $\delta
\hat\Gamma_k/\delta A^a_\mu(p)\times\delta\hat\Gamma_k/\delta K^a_\mu(-p)$ out
of the first three "classical" terms survives (note that $\hat\Gamma_k$ can be
assumed to be local due to the presence of the infra-red cutoff). Furthermore
$\hat\Gamma_k$ within the second factor of this term can be replaced by
$\hat\Gamma_0$, and the first factor is proportional to the gauge field mass
term mentioned above.

Together with the numerous terms with factors of $R^k_{\mu\nu}$ or $\tilde
R^k$ the modified ST identity (25) thus provides us with a one loop equation
for the gauge field mass term. In $d$ dimensions, and for an arbitrary cutoff
function $R^k$ in eqs. (11), this equation becomes
\begin{eqnarray}\label{29}  &m^2=&\frac{Ng^2}{16\pi^2}\int_{0}^{\infty}
dp^2(p^2)^{\frac{d}{2}-2}\frac{R^k(p^2)}{(1+R^k(p^2))^2}\Bigl\lbrace
\frac{11}{2}-d-\frac{5}
{d}+\alpha(1-\frac{1}{d})\nonumber\\
&&+\frac{p^2\partial_{p^2}R_k(p^2)}{(1+R^k(p^2))}
(\frac{7}{2}-\frac{6}{d})\Bigr\rbrace.\end{eqnarray}

Using $d=4$ and the form (13) of the cutoff function $R^k(p^2)$, one recovers
the result (28) for $m^2$ derived via the integration of the flow equations.
This constitutes a nontrivial check of our formalism; at the diagrammatic
level the contributions to eqs. (28) and (29) are entirely different. (Using
partial integrations under the $dp^2$ integral, the equivalence of the
different contributions can already be established in arbitrary dimensions
and for arbitrary cutoff functions. More details will follow in a separate
publication.)

Of course, the procedure underlying this consistency check required the
knowledge of a solution of the standard ST identities, eq. (25) for $k\to 0$,
in the form of the classical action. In the case of applications, which are
envisaged in the future, this knowledge is not available; $\hat\Gamma_k$ for
$k\to 0$ is the object to be computed in terms of $\hat\Gamma_k$ for
$k=\Lambda$ by integrating the flow equations.

To this end solutions to the modified ST identities (25) have to be found;
these can at least be constructed perturbatively along the lines discussed in
[5] for small coupling or large $\Lambda$ due to asymptotic freedom.

To conclude, the here presented modified ST identities provide an essential
step to treat non-abelian gauge theories with the help of flow
equations. These have already been proven to be a powerful and flexible method
elsewhere, but their applicability to gauge theories has been considered with
scepticism before.

The flow equations together with the modified ST identities might even be able
to serve as a nonperturbative definition of gauge theories; therefore,
however, the structure of the solutions for $k\to 0$ has to be studied in
detail. This problem is presently under investigation.

 It should be clear that additional matter can be included,
using the same methods and arguments discussed explicitly above, as long as
the theory remains asymptotically free. Finally, our results might also be of
help concerning the program of proofing renormalizability of gauge theories
within the framework of flow equations.

\end{document}